\documentclass[sigplan,9pt]{acmart}

\renewcommand\footnotetextcopyrightpermission[1]{} 
\acmConference[Middleware'19]{ACM/IFIP International  Middleware Conference}{9-13 December, 2019}{UC Davis, CA, USA}
\acmYear{2019}
 \acmISBN{978-x-xxxx-xxxx-x/YY/MM}
 \acmDOI{10.1145/nnnnnnn.nnnnnnn}
\startPage{1}

\setcopyright{acmlicensed}

\usepackage{booktabs}   
\usepackage{subcaption} 
 \usepackage{listings}                       
 \usepackage{color}
 \usepackage{tabularx}
 \usepackage{lscape}
 
\definecolor{lightgray}{rgb}{.9,.9,.9}
\definecolor{darkgray}{rgb}{.4,.4,.4}
\definecolor{purple}{rgb}{0.65, 0.12, 0.82}

\lstdefinelanguage{JavaScript}{
  keywords={typeof, new, true, false, catch, function, return, null, catch, switch, var, if, in, while, do, else, case, break},
  keywordstyle=\color{blue}\bfseries,
  ndkeywords={class, export, boolean, throw, implements, import, this},
  ndkeywordstyle=\color{darkgray}\bfseries,
  identifierstyle=\color{black},
  sensitive=false,
  comment=[l]{//},
  morecomment=[s]{/*}{*/},
  commentstyle=\color{purple}\ttfamily,
  stringstyle=\color{red}\ttfamily,
  morestring=[b]',
  morestring=[b]"
}

\usepackage{xparse}
\usepackage{algorithm}
\usepackage[noend]{algpseudocode}       
\NewDocumentCommand{\Event}{ O{Port} O{Event} O{} }{%
  \textit{#1}:\texttt{#2}$\langle$\textit{#3}$\rangle$
}

\algblockdefx[UponBlock]{Upon}{EndUpon}%
     {\textbf{upon}~}

\NewDocumentCommand{\Trigger}{O{\Event}}{\State \textbf{trigger}~#1}

\algblockdefx[ThreadBlock]{Thread}{EndThread}%
     {\textbf{thread}~}

\newcommand{\Wait}{\State \textbf{wait}~}                 
\algblockdefx[InterfaceBlock]{Interface}{EndInterface}%
     [2][\textit{name}]{\textbf{interface}~\texttt{#1}~(#2)}

\algloopdefx[ForAllBlock]{For}%
     [2][\textit{condition}]{\textbf{for~all}~#1~\textbf{in}~#2~\textbf{do}~}

\begin{document}

\title{Pando: Personal Volunteer Computing in Browsers}         


\author{Erick Lavoie, Laurie Hendren}
\affiliation{
  \institution{McGill University, Montreal, Canada}            
}
\email{erick.lavoie@mail.mcgill.ca}          
\email{hendren@cs.mcgill.ca} 


\author{Frederic Desprez}
\affiliation{
  \institution{INRIA Grenoble Rh\^one-Alpes}            
  \city{Grenoble}
  \country{France}                    
}
\email{Frederic.Desprez@inria.fr}          

\author{Miguel Correia}
\affiliation{
  \institution{INESC-ID}            
  \city{Lisboa}
  \country{Portugal}                    
}
\email{miguel.p.correia@tecnico.ulisboa.pt}          

\begin{abstract}
 The large penetration and continued growth in ownership of personal electronic devices represents a freely available and largely untapped source of computing power. To leverage those, we present Pando, a new volunteer computing tool based on a declarative concurrent programming model and implemented using JavaScript, WebRTC, and WebSockets. This tool enables a dynamically varying number of failure-prone personal devices contributed by volunteers to parallelize the application of a function on a stream of values, by using the devices' browsers. We show that Pando can provide throughput improvements compared to a single personal device, on a variety of compute-bound applications including animation rendering and image processing. We also show the flexibility of our approach by deploying Pando on personal devices connected over a local network, on Grid5000, a French-wide computing grid in a virtual private network, and seven PlanetLab nodes distributed in a wide area network over Europe.
\end{abstract}



\maketitle

\section{Introduction}
More than 1.5 billion smartphones were sold in the world in 2018~\cite{gartner2018cellphonesold} and the computing power of the highest-end devices today rivals that of desktops and laptops~\cite{herrera18webassembly}. They collectively represent an \textit{immense source of largely untapped computing power}.  

While the latest developments in distributed computing have had tremendous impact in industry and elsewhere, the major paradigms that sustained those developments have led to designs with barriers that limit the utilization of personal devices for distributed computing~\cite{lavoie2019pvc-cf}: access to cloud platforms require \textit{financial instruments}, such as a bank account or a credit card; access to grid platforms require \textit{administrative permissions}; and the deployment of the most popular volunteer computing platform, BOINC~\cite{anderson2004boinc}, requires a significant \textit{technical effort} because it has been designed for long-running large-scale research projects with contributors that are anonymous and potentially malicious. In a sense, the underlying problem is socio-technical: \textit{we do not have technical solutions that can leverage, in a seamless way, the abundance of computing power we collectively already possess}.

Recently, we have proposed \textit{personal} volunteer computing~\cite{lavoie2019pvc-cf} to address this problem. In contrast to volunteer computing, the approach focuses on the development of personal tools, for personal projects, that leverage the computing capabilities of personal devices owned by users and their friends, family, and colleagues. However, a comprehensive description of an example tool that could do so had yet to be published.

In this paper, we therefore present Pando, a new tool that can leverage a dynamically varying number of failure-prone personal devices contributed by volunteers, to parallelize the application of a function on a stream of values, by using the devices' browsers. Pando is based on a \textit{declarative concurrent} programming paradigm~\cite{van2004concepts} which greatly simplifies reasoning about concurrent processes: it abstracts the non-determinism in the execution by making it non-observable. This paradigm has already enjoyed great practical successes with the popular MapReduce~\cite{dean2008mapreduce} and Unix pipelining~\cite{Kernighan1983} programming models. We show for the first time it is also effective in personal volunteer computing tools.

Pando abstracts distribution but otherwise relies on existing toolchains: programmers define the function to distribute and the modules it depends on following the current JavaScript programming idioms, and users can easily combine Pando in Unix pipelines. Deployment on volunteers' devices simply requires opening, in their browser, a URL provided by Pando on startup. Devices may join or quit at any time and Pando will transparently handle the changes. We present both the high-level design principles that guided the design and a concrete working implementation, itself organized around the pull-stream design pattern and based on JavaScript~\cite{javascript}, WebSockets~\cite{websocket}, and WebRTC~\cite{webrtc} to enable its execution inside browsers. The implementation of Pando is open source~\cite{pando-repository}. Compared to other volunteer computing tools, we conceived Pando as a personal tool for quick and easy deployment rather than as a long-running server process. We also avoided the use of a database for tracking the status of inputs and leveraged the heartbeat mechanism of WebSockets and WebRTC to simplify the implementation of fault-tolerance.	

The programming model of Pando corresponds to a streaming version of the functional \textit{map} operation that supports a dynamic number of devices, without an \textit{a priori} limit on their number. It reads new inputs only when computing resources are available for processing and tolerates failures in which devices suddenly disconnect, either intentionally or by crashing. To maximize throughput, faster devices receive more inputs and only a single copy of an input is submitted for processing at a time. Those properties are encapsulated in a reusable abstraction, StreamLender, that is independent of the communication protocols and input-output libraries we used for the implementation. StreamLender requires only higher-order functions for its implementation, making it portable to many popular programming languages of today. We describe the key aspects of the implementation of StreamLender. We also provide the JavaScript implementation used by Pando as a reusable JavaScript library~\cite{pull-lendstream-implementation}. To the best of our knowledge, StreamLender is the first articulation of those properties in a reusable abstraction for distributed stream processing.

We have applied Pando to seven compute-bound applications, including crypto-currency mining, crowd computing, machine learning hyper-parameter optimization, and open data processing in combination with other peer-to-peer data distribution protocols. This effort has highlighted the suitability of Pando's programming model to common processing pipelines but also the possibility of integrating Pando as a component in applications with more complex feedback loops, e.g. when performing synchronous parallel search or handling failures in external data distribution protocols.

We have deployed Pando on personal devices in a local-area network on our personal collection of devices, on Grid5000~\cite{grid5000}, a French-wide computing grid that regroups multiple clusters of computing nodes in a virtual private network (VPN) similar to the computing resources available to a large organization, as well as on seven PlanetLab computing nodes contributed by various organizations throughout Europe, connected over a wide-area network (WAN). By batching inputs for distribution, the network latency could be hidden, and we achieved overall throughput higher than on a single personal device, regardless of the position of the computing devices in the network. This shows that Pando can take advantage of both local and remote devices. To the best of our knowledge, it is the first time a tool for volunteer computing has been shown to be easily deployable in all three settings. Moreover, the comparison between the performance of recent personal devices and high-end servers shows that 2-5 cores on a personal device can outperform a core on a high-end server, highlighting the competitive opportunity offered by personal devices contributed by volunteers.

The rest of this paper is organized as follows. We present the overall design of Pando in Section~\ref{Section:Design}. We provide the key properties and behaviour of the StreamLender abstraction in Section~\ref{Section:StreamLender}. We present the different applications in Section~\ref{Section:Applications} and evaluate the benefits and limitations of parallelizing them in real-world deployments in Section~\ref{Section:Evaluation}. We compare the specificities of our design to related work in Section~\ref{Section:RelatedWork}.  We conclude with a brief recapitulation of the paper and future work in Section~\ref{Section:Conclusion}.

\section{Pando}
\label{Section:Design}

Pando is the first tool explicitly designed for the purpose of personal volunteer computing. We first explain how to use it and its concrete benefits using one of our supported application (Section~\ref{Section:UsageExample}). We then articulate the design principles that enable those benefits (Section~\ref{Section:DesignPrinciples}). We continue with a more detailed explanation of Pando's programming model (Section~\ref{Section:ProgrammingModel}) and finally present an overview of how it is implemented in a concrete system (Section~\ref{Section:Implementation}).

\subsection{Usage Example}
\label{Section:UsageExample}

\begin{figure}[htbp]
    \centering
    \begin{subfigure}[t]{0.11\textwidth}
        \includegraphics[width=\textwidth]{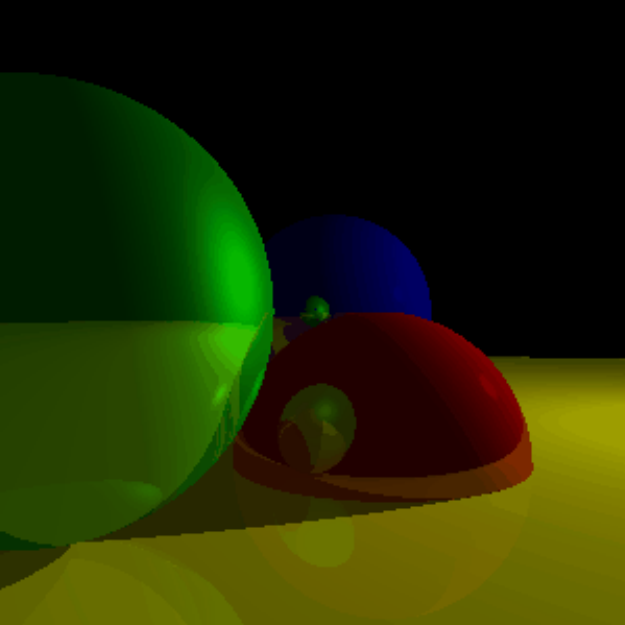}
    \end{subfigure}
    ~ 
    \begin{subfigure}[t]{0.11\textwidth}
        \includegraphics[width=\textwidth]{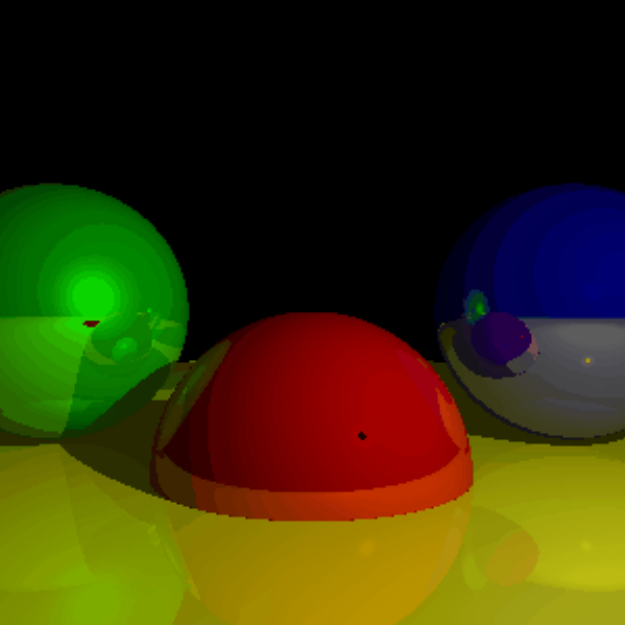}
    \end{subfigure}
     ~ 
    \begin{subfigure}[t]{0.11\textwidth}
        \includegraphics[width=\textwidth]{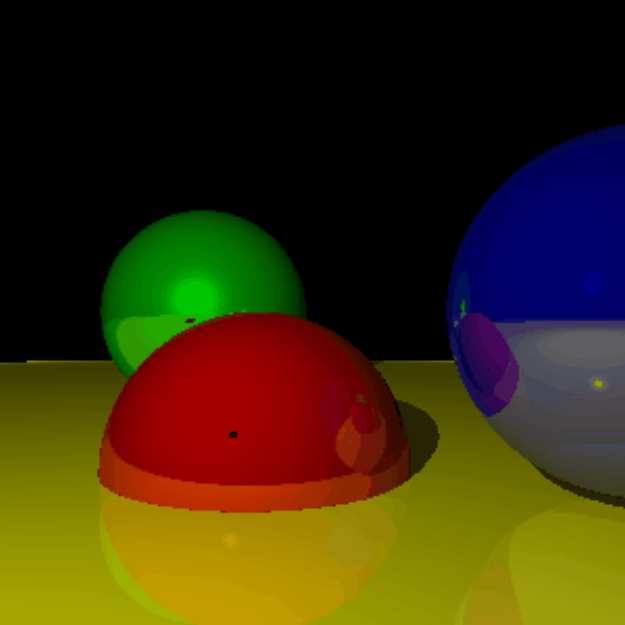}
    \end{subfigure}
    \caption[Rotation animation]{Rotation animation around a 3D scene.}
    \label{Figure:AnimationFrames}
\end{figure}

Suppose a user is working on a personal project involving an animation, as shown in Figure~\ref{Figure:AnimationFrames}, and the rendering uses \textit{raytracing}~\cite{whitted1980raytracing}, which is computationally expensive. To accelerate the rendering of the entire animation, they want to \textit{parallelize} the rendering of individual frames, while still obtaining them in the \textit{correct order}.

If this were a professional project, our user could rely on professional solutions~\cite{zincrender,deadline}. However, these are often too expensive for personal projects and do not easily leverage the computing power of devices users already own. Instead, they can use Pando through a \textit{simple programming interface} and a \textit{quick deployment} solution.

\subsubsection{Programming Interface}

Pando's distribution of computation is organized around a \textit{processing function} which is applied to a stream of \textit{input values} to produce a stream of \textit{outputs}. In this particular example, the processing function performs the \textit{raytracing of the scene} from a particular \textit{camera position} and outputs an \textit{array of pixels}. The animation consists in a sequence of positions of the camera rotating around the scene. 

Pando's implementation parallelizes the execution of code in JavaScript by using the Web browsers of personal devices. To leverage those capabilities, a user writes a minimal amount of glue code to make the processing function compatible with Pando's interface, as illustrated in Figure~\ref{Listing:ProgrammingInterface}. In this example, the raytracing operation is provided by an external library, taken unmodified from the Web, which is first imported. Then a processing function using the required library is exposed on the module with the \texttt{'/pando/1.0.0'} property, which indicates it is intended for the first version of the Pando protocol. The function takes two inputs: \texttt{cameraPos}, the camera position for the current frame and \texttt{cb}, a callback to return the result. The body of the function first converts the camera position, which was received as a string, into a float value, then renders the scene. The \texttt{pixels} of the rendered image are then saved in a buffer, compressed with \texttt{gzip}, and output as a base64 encoded string~\cite{base64}, which simplifies its transmission on the network.\footnote{Those last three operations take a negligible amount of time compared to rendering.} The result is then returned to Pando through the callback \texttt{cb}. In case an error occurred in any of those steps, an error is caught then returned through the same callback.

\begin{figure}[htbp]
\begin{lstlisting}
// Import existing function
var render = require('raytracer')
 // Import compressing module
var zlib = require('zlib')
module.exports['/pando/1.0.0'] = function(cameraPos, cb) {
    try {
      var pixels = render(parseFloat(cameraPos))
      cb(null, zlib.gzipSync(new Buffer(pixels)).toString('base64'))
    } catch (err) {
      cb(err)
    }
}
\end{lstlisting}
\caption[JavaScript programming interface example.]{JavaScript programming interface example for rendering with raytracing.}
\label{Listing:ProgrammingInterface}
\end{figure}

The glue code should then be saved in a file, \texttt{render.js} in this example, and all library dependencies should be accessible using the Node Package Manager (NPM) conventions~\cite{npm}, typically in a \texttt{node\_modules} sub-directory.  Pando will automatically bundle all the dependencies on startup and adapt the code for the browser context by internally using \texttt{browserify}~\cite{browserify}.

Pando is compatible with the Unix standard process interface, i.e. it can either receive its inputs on the \textit{standard input} or as command-line arguments and it produces outputs on the \textit{standard output}.  In Figure~\ref{Listing:UnixProgrammingInterface}, we connect Pando with other tools using bash scripting. The camera positions are provided as strings on the standard input by \texttt{generate-angles.js}, the rendered images are produced on the standard output as strings by Pando, and the assembly of the frames into a GIF animation is done by \texttt{gif-encoder.js}. All tools in the sequence are connected through Unix streams using the pipe operator ('\texttt{|}'). Pando could also be scripted from any other programming environment that supports the creation of Unix processes; the creation of inputs and the post-processing of outputs therefore need not be in JavaScript.

\begin{figure}[htbp]
\begin{lstlisting}[language=Bash]
$ ./generate-angles.js | pando render.js --stdin | ./gif-encoder.js
Serving volunteer code at http://10.10.14.119:5000
\end{lstlisting}
\caption[Unix programming interface example.]{Unix programming interface example for rendering inputs and processing outputs. After starting, Pando lists the URL necessary for deployment on the standard error.}
\label{Listing:UnixProgrammingInterface}
\end{figure}
 
\subsubsection{Deployment}

\begin{figure*}[htbp]
    \centering
       \subcaptionbox{\label{Figure:ExecutionExampleStart}  Initial state.}{\includegraphics[width=0.18\textwidth]{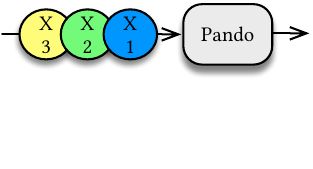}}
       \subcaptionbox{\label{Figure:ExecutionExampleTabletJoins} A tablet joined.}{\includegraphics[width=0.155\textwidth]{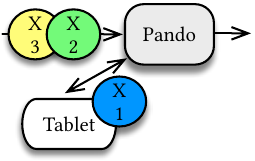}}
       \subcaptionbox{\label{Figure:ExecutionExampleTabletDone1} Tablet rendered $x_1$.}{\includegraphics[width=0.18\textwidth]{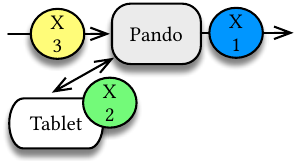}}
       \subcaptionbox{\label{Figure:ExecutionExamplePhoneJoins} A phone joined.}{\includegraphics[width=0.18\textwidth]{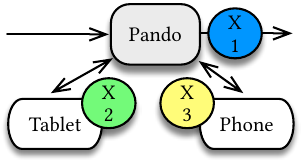}}
       \subcaptionbox{\label{Figure:ExecutionExamplePhoneDone3} Phone rendered $x_3$.}{\includegraphics[width=0.18\textwidth]{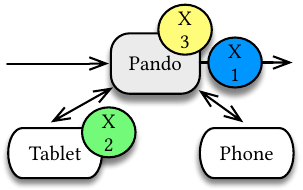}}
\end{figure*}

\addtocounter{figure}{-1}
\begin{figure}[htbp]
      \subcaptionbox*{\setcounter{subfigure}{5}}{}
       \subcaptionbox{\label{Figure:ExecutionExampleTabletCrash} Tablet crashed.}{\includegraphics[width=0.155\textwidth]{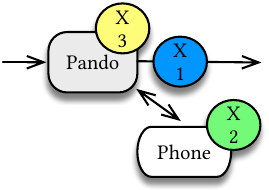}}
       \subcaptionbox{\label{Figure:ExecutionExamplePhoneDone2} Phone rendered $x_2$. Processing is over.}{\includegraphics[width=0.18\textwidth]{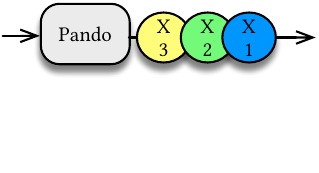}}    
	 \caption[Deployment]{Deployment example.}
	\label{Figure:Deployment}
\end{figure}

A user deploys Pando by starting it on the command-line\footnote{After installing, ex: \texttt{npm install --global pando-computing}~\cite{pando-handbook}.}, as illustrated in Figure~\ref{Listing:UnixProgrammingInterface}. Then they should wait for URL messages to appear. When displayed, those messages indicate that Pando is ready for other devices to join.

A user then opens the URL in the browser of its personal devices. Upon joining, additional devices will process individual frames in parallel. In one possible example execution, illustrated in Figure~\ref{Figure:Deployment}, a tablet joins after the volunteer URL has been opened, then renders an image, then a faster phone joins, also renders an image, then the tablet crashes, and the phone takes over for the missing image. Communications happen over a choice of WebRTC~\cite{webrtc}, a recent peer-to-peer protocol for browsers, or WebSocket~\cite{websocket}.

A user can invite friends to add their devices, even if they are outside the local network. To do so, the user deploys a small micro-server we built for Pando~\cite{pando-server} on a platform that provides a public IP address, such as Heroku~\cite{heroku}. Being publicly accessible, the URL can then be shared to friends on existing social media. After opening the URL, a WebRTC connection will directly connect joining devices.

 As illustrated in this deployment example, Pando \textit{dynamically scaled} to accommodate the number of participating devices and \textit{gracefully tolerated failures} with no particular programming effort from the user beyond specifying a function to process a single value. Moreover, the user did not need to (1) buy new devices, (2) create an account or obtain administrative permissions, (3) use financial instruments, (4) accommodate device specificities, or (5) wait for resources to be freed. The user could also (1) combine Pando with existing Unix tools, (2) use social media to request for help, and (3) know their data has only been shared between trusted devices.

\subsection{Design Principles}
\label{Section:DesignPrinciples}

The previous usage example provided significant benefits because we designed Pando around the following design principles (DPs), which we derived from the limitations of previous approaches~\cite{lavoie2019pvc-cf}.

\textit{Specific deployment} (DP1): the deployment of the tool that connects the different volunteers is specific to: (1) a single project, (2) a single known user with an existing social presence, either through the contacts of volunteers, or an identity in a social platform, and (3) the lifetime of the corresponding tasks, after which it shuts down.

\textit{Compatible with a wide variety of existing personal devices} (DP2): the tool should leverage desktops, laptops, tablets, phones, embedded devices, and personal appliances that people already own.

\textit{Easy to program} (DP3): the implementation of tasks should be done with a minimum of programming effort for use in a distributed setting. Ideally, it should be as easy to program in a distributed setting as in a local one.

\textit{Quick to deploy} (DP4): the tool should require little installation effort, should start processing quickly after launch, and then should dynamically scale up to benefit from help obtained from friends' devices.

\textit{Composable and modular} (DP5): the tool should focus on coordinating contributing volunteers' devices but otherwise should rely on other tools and technologies for the rest of the needs of users. The core abstractions used in particular tools should be applicable to other uses. Tools should also combine with high-performance libraries, when available, to leverage the latest results of parallelism research without making the tools themselves more complicated.

\subsection{Programming Model}
\label{Section:ProgrammingModel}

In effect, Pando's programming model corresponds to a streaming version of the functional \textit{map} operation: Pando applies a function $f$ on a series of input values $x_i$ to obtain a serie of results $f(x_i)$. Its implementation is free to process inputs in any order but outputs results in the order of their corresponding inputs.

We chose a streaming programming model because it is simple to program (DP3) yet powerful enough to coordinate the usage of multiple devices in parallel (DP2). The reason is that it belongs to the \textit{declarative concurrency} paradigm~\cite{van2004concepts} which \textit{abstracts the non-determinism of executions by making it non-observable to the programmer}. In other words, a declarative concurrent program outputs the same result regardless of the order in which the various threads that compose the execution complete their tasks. That makes Pando as simple to program in a sequential setting with a single participating processor as for a parallel case with dozens. While it is implied by the definition of the \textit{map} operation, it is worth noting that the \textit{ordering} of outputs is important to preserve the \textit{declarative concurrency} property; otherwise the relative speed of processors could influence the order of the results and make the non-determinism observable. Note also that an implementation of $f$ may have side-effects, such as pulling data and transferring back results to a server, while maintaining the benefits of \textit{declarative concurrency}. In this case however, it is the responsibility of the programmer to ensure that the order of side-effects does not matter.

We initially chose the \textit{streaming map} programming model because it fits more problems than the \textit{bag-of-tasks} model of typical volunteer computing problems, which usually have independent inputs with no ordering requirement. Some applications however, such as the sequence of images that compose the animation of our previous example (Section~\ref{Section:UsageExample}), do require a particular order. Problems with \textit{unordered} inputs can be reduced to a streaming version simply by incrementally traversing the values in an arbitrary order, making the streaming model more general. The streaming version also enables working with an \textit{infinite} number of values and applications requiring \textit{feedback loops} (Section~\ref{Section:Applications}).

We also chose a number of additional distributed properties for Pando to make it easy to program (DP3) and fast to deploy (DP4). 

First, participating devices may join \textit{dynamically}, at any time during execution. Pando's computing power will grow automatically. This removes the overhead of registering computing resources in advance and simplifies scaling for quick deployment. 

Second, The potential number of participating devices is \textit{unbounded}. Pando strives to provide the illusion of infinite scalability so its actual performance grows automatically as users adopt new devices with more capabilities. 

Third, Pando is also \textit{lazy}: i.e. it reads inputs only when computing resources become available. This adjusts the flow of values  to the available computing power to avoid overloading Pando's memory with pending values. It also makes the implementation compatible with infinite streams with no additional effort. Users get support for laziness with no additional programming effort.

Last, Pando also \textit{tolerates failures} of participating devices, making those failures transparent to the programmer. We chose a \textit{crash-stop} failure mode\footnote{Failure modes can  range from \textit{crash-stop}, in which a process follows its instructions then may crash and stop sending messages forever, passing by \textit{crash-recovery}, in which a process may fail then recover and try participating again, to \textit{byzantine}, in which a process may deviate arbitrarily from its instructions including intentionally sending messages to hamper progress.}, in which participating devices will always faithfully carry their assigned task without deviating from their prescribed behaviour until they either suddenly crash or disconnect. This model corresponds to failures in which a browser tab, that executes computations, is suddenly closed or to a loss of network connectivity. In the presence of such failures, Pando guarantees \textit{liveness}: once an input $x_i$ has been read, if there are active participating devices, Pando will eventually provide $f(x_i)$.

The crash-stop failures of participating devices can be detected because we assume a \textit{partially synchronous} execution\footnote{Timing assumptions may range from \textit{fully synchronous}, in which there is an upper time bound on message delivery, passing by \textit{partially synchronous}~\cite{dwork1988partialsynchrony}, in which there is a time bound on delivery that it will apply only \textit{eventually} after an unknown delay, and culminating in \textit{asynchronous}, in which there are no time bound on delivery.}: most of the time, messages will be delivered within a specified \textit{time bound}. This corresponds to the ability of communication channels such as TCP~\cite{tcp} and WebRTC~\cite{webrtc} to suspect failures by failing to receive the acknowledgment of a \textit{heartbeat} message within a time bound.

In terms of performance goals, we decided to focus on maximizing \textit{throughput} with the additional following two properties. Pando distributes values to participating devices \textit{conservatively}: a value is sent to at most one device for processing. The device will either produce a result or will crash, in which case the value will be sent to another device. This ensures participating devices process a maximum number of values simultaneously. Moreover, the rate at which values are submitted to participating devices \textit{adapts} to their processing speed. Devices with a faster processing speed will receive more values to process, maximizing resource utilization.

This combination of programming model properties, summarized in Table~\ref{Table:PandoProperties}, provides a powerful yet easy-to-use programming model as shown by the breath of applications supported (Section~\ref{Section:Applications}). 

\begin{table}[htbp]
\centering
\begin{tabular}{l l}
\toprule
\textbf{Streaming Map}            &  $x_1,x_2,... \rightarrow f(x_1), f(x_2), ...$. \\
\textbf{Ordered}                       &  Outputs provided in order. \\
\hline
\textbf{Dynamic}                      &  New devices may join any time. \\
\textbf{Unbounded}                  & No \textit{a priori} limit on participants nb. \\
\textbf{Lazy}                             & Inputs read when resources are avail.  \\
\textbf{Fault-tolerant}               &  \textit{Crash-stop} failures  are tolerated. \\
\hline
\textbf{Conservative}               & A single copy submitted at a time. \\
\textbf{Adaptive}                   & Faster devices receive more inputs. \\
\bottomrule
\end{tabular}
\caption{Summary of the programming model properties.}
\label{Table:PandoProperties}
\end{table}

\subsection{Implementation Overview}
\label{Section:Implementation}

Our implementation was first based on our choice between available Web technologies (Section~\ref{Section:ImplementationChoices}). We then organized it around a declarative concurrent paradigm to simplify both its usage and implementation effort (Section~\ref{Section:PullStream}). We finally designed a reusable architecture by decomposing it into modules and communication technologies (Section~\ref{Section:Architecture}).

\subsubsection{Technology Choices}
\label{Section:ImplementationChoices}

We based our implementation on Web technologies for a number of reasons. First, they are compatible with a wide number of personal devices, from smartphones and embedded devices to tablets, laptop, and desktops computers (DP2). Second, virtual machines in modern browsers execute numerical applications in JavaScript at a speed within a factor of 3 of equivalent numerical code written in C \cite{Khan:2014,herrera18webassembly}. A large variety of native applications, as represented by the SPEC CPU2006 and CPU2017 benchmarks and originally written in C for Unix systems, can also be executed in browsers supporting WebAssembly~\cite{webassembly} without modification to the original source code by using Browsix-WASM~\cite{jangda2019browsix-wasm}: the applications then run with an average slowdown of only 45\% to 55\% and peak slowdown of 2.5x compared to a native execution. In either case, the level of performance is sufficiently close to C to benefit from executing tasks inside multiple parallel Web pages. Third, browsers also provide a security sandbox that prevents code executing within a web page from tampering with the host operating system.  Fourth, \textit{WebRTC} \cite{webrtc}, enables direct communication between browsers, in many cases even in the presence of Network Address Translation (NAT), which removes the need for a server to relay all communications between the tool and the volunteers' devices. Fifth, links shared on social media platforms enable their users to quickly mobilize their social networks. Sixth, both \textit{WebSocket}~\cite{websocket} and \textit{WebRTC}~\cite{webrtc} provide heartbeats to detect disconnections. 

\subsubsection{Declarative Concurrency With Pull-Streams}
\label{Section:PullStream}

Pando provides a \textit{declarative concurrent} abstraction~\cite{van2004concepts} of the parallel execution of the different participating processors (Section~\ref{Section:ProgrammingModel}).  Mainstream languages, such as JavaScript, have not yet integrated features that make that style of programming widely accessible. We therefore instead based our design and implementation on the pull-stream design pattern~\cite{pull-streams}, a functional code pattern that enables streaming modules to be built by following a simple callback protocol. It only requires support for higher-order functions from the base language. Implementations of abstractions built by following the pattern should therefore be straight-forward to port to many programming languages of today.

The pull-stream design pattern has originally been proposed by Dominic Tarr~\cite{pull-streams} as a simpler alternative to Node.js streams, that were plagued with design issues that had to be maintained for backward-compatibility. A community has grown around the pattern and more than a hundred modules have been contributed~\cite{pull-stream-modules}. 

\begin{figure}[htbp]
    \begin{center}
    \includegraphics[width=0.40\textwidth]{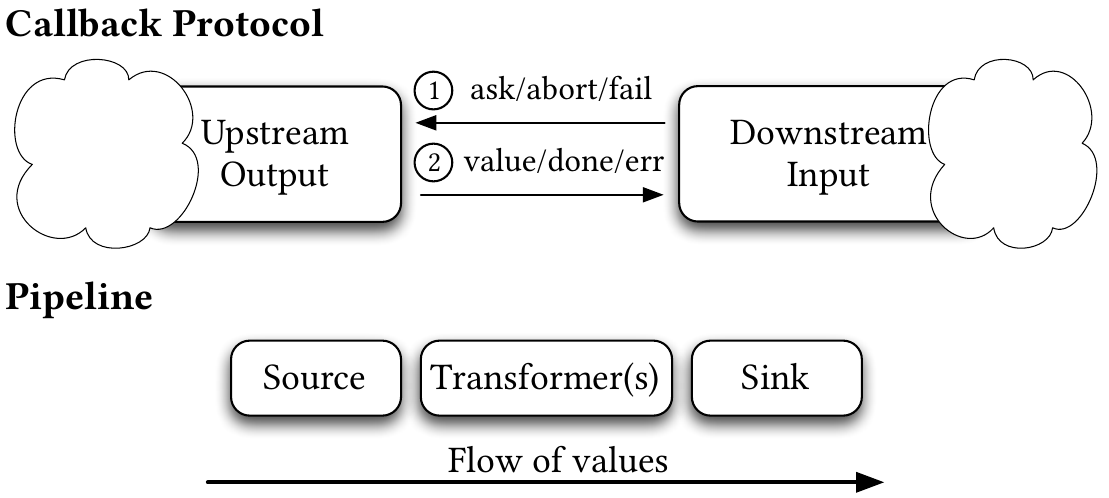}
    \end{center}
    \caption[Pull-stream design pattern.]{\label{Figure:PullStream} Pull-stream design pattern: callback protocol on top and pipeline of composable modules at the bottom.}
\end{figure}

Perhaps, the simplest example of pull-stream modules is a source that lazily counts from $1$ to $n$, connected to a sink that consumes all values and then stops, as illustrated in Figure~\ref{Listing:PullStreamExample}. The callback protocol essentially consists in a request followed by an answer. The request may be used to ask for a value, abort the stream normally, or fail because of an error. Symmetrically, the answer may then produce a value, signify the end of the stream, or stop because of an error. A module may also both consume and produce values, in which case it can be used between a source and a sink. This is illustrated in Figure~\ref{Figure:PullStream}. 

\begin{figure}[htbp]
\begin{lstlisting}
function source (n) {
  var i = 1
  return function output (abort, cb) {
    if (abort) 
      return cb(abort, undefined)
    else if (i<=n)
      return cb(false, i++)
    else
      return cb(true, undefined)
  }
}
function sink (request) {
  request(false, function answer (done, v) {
    if (done) return
    else request(false, answer)
  })  
}
sink(source(10))
var pull = require('pull-stream')
pull(source(10), sink) // equivalent to line 20
\end{lstlisting}
\caption{Pull-stream example.}
\label{Listing:PullStreamExample}
\end{figure}

While the pattern does not simplify the task of implementing pull-stream modules, once implemented, the modules provide clear semantics and are easy to combine because they can provide declarative concurrent abstractions. Using the pull-stream design pattern therefore makes the rest of the implementation of Pando easier.

\subsubsection{Architecture}
\label{Section:Architecture}

\begin{figure}[htbp]
    \begin{center}
    \includegraphics[width=0.46\textwidth]{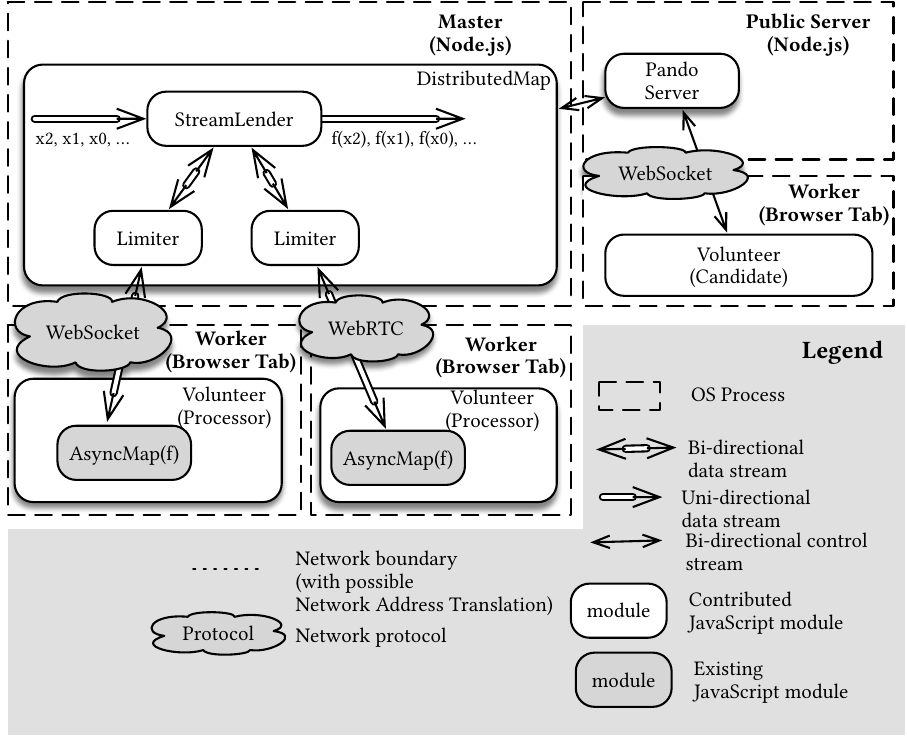}
    \end{center}
    \caption{\label{Figure:Architecture} Architecture of Pando.}
\end{figure}

The core modules of Pando and the way they are connected is illustrated in Figure~\ref{Figure:Architecture}. They work together to implement a \textit{distributed map} that processes a stream of values $x_i$ with a function $f$. Our implementation uses Node.js but could also work as a hosted Web application. Deployment consists in executing the tool on the command-line, which starts the Master process. HTTP connections from volunteers' devices may then be made directly to the Master, if on the same local area network (not shown), or through a Public Server, if direct connectivity is not possible. The HTTP connection is used to obtain the Worker code including the $f$ function and eventually establish either a WebSocket~\cite{websocket} or WebRTC~\cite{webrtc} connection. The bootstrap of the WebRTC connection, which requires \textit{signalling} of possible connection endpoints between peers, is done through a Public Server using a separate WebSocket connection. That connection closes after the WebRTC connection is established. Since signalling requires little resources, the Public Server could be executed on a small personal server such as a Raspberry Pi board~\cite{raspberry-pi} or the free tier of a cloud such as Heroku~\cite{heroku}.

The pull-stream abstractions we designed and reused are shown as modules within the different processes, respectively in white and grey. The core coordination is performed by our novel \textit{StreamLender} abstraction (Section~\ref{Section:StreamLender}), which creates multiple concurrent bi-directional sub-streams, one for each worker. A sub-stream continuously borrows values from the input of StreamLender and return results that are eventually returned on its output. The sub-streams are dynamically created as Workers join. We use existing libraries that expose WebRTC and WebSocket channels as pull-streams. Since their implementation \textit{eagerly reads all available values} on the sending side, we bound the total number of values that can be borrowed using our new \textit{Limiter} module: initially a bounded number of inputs is let through until the limit is reached, then for each new result that comes in a new input is allowed. With a large enough limit, data transfers in both directions therefore happen in parallel with the computations and can hide transmission latency. The limit can be parameterized using an argument passed to Pando on startup. The actual processing of values is done inside Workers using the existing \textit{AsyncMap}~\cite{pull-stream-modules} module that applies the function $f$ on the different inputs. 

Pando trivially enables parallel processing on multicore architectures on a single machine while enabling dynamically scaling up to other devices if necessary, making the tool useful in many contexts. Our design should also work with other technology choices, which could be mandated because users require specific libraries and technologies that are not available for the Web yet. For example, users may depend on specific numerical libraries available in Python/Numpy, MATLAB, or R. In that case, it should be straightforward to adapt the design by relying on TCP for communication and porting our modules to a different language.

\subsection{Applicability}

The design and architecture of Pando are tailored to its application context: the acceleration of personal workloads with personal devices. Most of these workloads do not require strong timing guarantees, as could occur in real-time processing of sensor data or financial transactions for example. Moreover, a user has direct control over many or most of the personal devices that are used for computation: faults that may happen are the result of a user disconnecting a device accidentally or because it is not contributing significantly to the overall throughput. Fault-tolerance makes the tool more convenient to use but is not critical for efficient execution. Finally, it is easy to protect a Pando deployment against a denial-of-service attack because there is no long-running publicly accessible platform to target: an attacker needs to know \textit{when} a deployment happens, in addition to \textit{where}. It is also always possible to only deploy Pando behind a virtual private network for additional guarantees. The design of Pando therefore leverages the application context to simplify its implementation and therefore occupies a different part of the design space than many other distributed computing platforms.

\section{StreamLender}
\label{Section:StreamLender}

\begin{figure}[htbp]
    \begin{center}
    \includegraphics[width=0.26\textwidth]{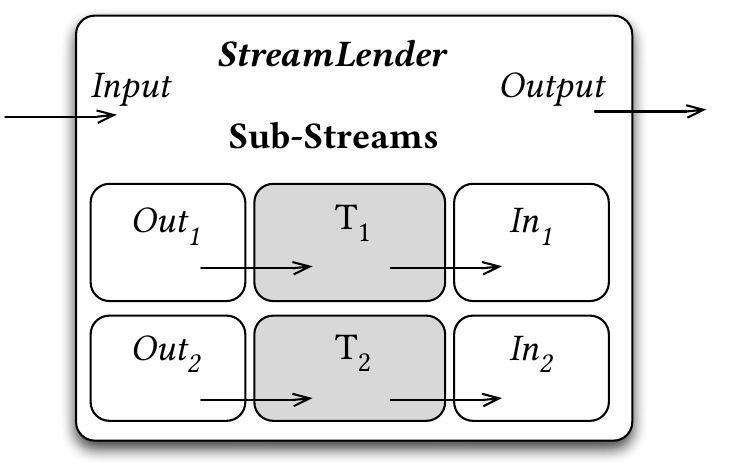}
    \end{center}
    \caption{\label{Figure:StreamLender} StreamLender and its sub-streams. External transformer(s) modules connected to the sub-streams are greyed. They represent modules such as the Limiter of Figure~\ref{Figure:Architecture}.}
\end{figure}

StreamLender is our novel abstraction that splits an input stream into multiple concurrent sub-streams and then merges back the results in a single output stream. The actual processing of the values is done using other transformer modules, as illustrated in Figure~\ref{Figure:StreamLender}. We provide a usage example in Figure~\ref{Listing:StreamLenderUsage}.

\begin{figure}[htbp]
\begin{lstlisting}
var pull = require('pull-stream')
// StreamLender
var lender = require('pull-lend-stream')
var limit = require('pull-limit') // Limiter
pull(
  pull.count(10),
  lender,
  pull.drain()
)
var duplex = ... // On webrtc connection opened
lender.lendStream(function (err, subStream)) {
  if (err) return
  pull(
    subStream.source, // output
    limit(duplex),
    subStream.sink     // input
  )
})
\end{lstlisting}
\caption{StreamLender usage example.}
\label{Listing:StreamLenderUsage}
\end{figure}

StreamLender encapsulates the \textit{streaming}, \textit{ordered}, \textit{dynamic}, \textit{fault-tolerant}, \textit{conservative}, and \textit{adaptive} properties of Pando's programming model (Section~\ref{Section:ProgrammingModel}), independently of a particular communication protocol or other input-output libraries. To the best of our knowledge, StreamLender is the first articulation of those properties in a reusable abstraction for distributed stream processing. 

The complete and tested JavaScript implementation that we built and used in Pando is available as an independent pull-stream module~\cite{pull-lendstream-implementation}. The synchronization of events happening through callbacks initiated by multiple concurrent streams was tricky to correctly implement and is rather cumbersome to decipher through the source code. We therefore derived a more readable pseudo-code version that uses explicit waiting primitives and events that correspond to the invocation of callbacks to help reimplementations, available in an extended version of this paper~\cite{lavoie2019thesis}. As a sample, Algorithm~\ref{Algorithm:StreamLenderImplementation} shows how the requests made on a sub-stream output are answered, either with a value from another sub-stream that failed, a new value requested on the StreamLender $Input$, or a $done$ if no more values are left to process. The ordering and synchronization of outputs is simply solved with a blocking queue that waits for the result at the next index in the stream to arrive.

\begin{algorithm}[h]
   \caption{Sub-stream output ask request.}
   \label{Algorithm:StreamLenderImplementation}
   \begin{algorithmic}[1]
     \Upon \Event[Out$_i$][ask][]
       \If{$failed \neq \emptyset$}
          \State \Call{answerWithFailedValue}{$Out_i$}
       \ElsIf{$Input$ has terminated ($done$ or $err$)	} 
           \State \Call{waitOnOthers}{$Out_i$}
       \Else \Comment{Lazily read a new value}
           \Trigger[\Event[Input][ask]]
           \Wait $Input$ answer
           \If{$answer =~$\Event[Input][value][v]}
               \State remember $v$
               \Trigger[\Event[$Out_i$][value][$v$]]
           \Else
                \State \Call{WaitOnOthers}{$Out_i$}
           \EndIf
       \EndIf
     \EndUpon
     
     \Procedure{answerWithFailedValue}{$Out_i$}
     	     \State let $v$ be the oldest value of \textit{failed}
              \State remember $v$
              \State $\textit{failed} \leftarrow \textit{failed} \backslash  \{v\}$
              \Trigger[\Event[Out$_i$][value][$v$]] 
     \EndProcedure
     
     \Procedure{waitOnOthers}{$Out_i$}
     	\Wait until last result received \textbf{or} $\textit{failed} \neq \emptyset$
	\If{last result received}
	    \Trigger[\Event[$Out_i$][done]]
	\Else
             \State \Call{answerWithFailedValue}{$Out_i$}
	\EndIf
     \EndProcedure
    \end{algorithmic}
\end{algorithm}

\section{Applications}
\label{Section:Applications}

Pando can be applied to a wide range of applications. In this section, we present some examples  according to their dataflow pattern, i.e. how data flows between Pando and other tools and protocols. We implemented each application using components built as separate Unix tools but the same components could be implemented as pull-stream modules and combined into a single application as well, either as a standalone webpage or a smartphone application. We summarize key aspects of each application.

\subsection{Pipeline Processing}
\label{Section:Pipeline}

\begin{figure}[htbp]
    \begin{center}
    \includegraphics[width=0.36\textwidth]{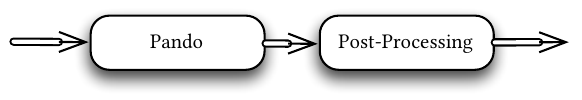}
    \end{center}
\begin{tabular}{llll}
\textbf{App.}  & \textbf{Inputs}           & \textbf{Pando}                    & \textbf{Post}   \\
\hline
Collatz           & Ints         & Nb of steps & Max            \\
Raytrace     & Camera pos.          & Raytracing                    & Anim. gif \\
Arxiv  & Meta-info          & Human tagging            & None                       \\
SL test & RNG seeds  & Rand. exec.   & Monitor fail.           \\
ML agent       & Hyperparams & Simulation           & None     \\ 
Img proc. & Landsat-8 imgs & Blur filter & None \\   
(http) & & & \\            
\end{tabular}
    \caption{\label{Figure:PipelineProcessing} Pipeline processing dataflow and examples.}
\end{figure}

\textit{Pipeline processing} is a sequence of independent processing stages applied to a stream of inputs, as illustrated in Figure~\ref{Figure:PipelineProcessing}. Traditional \textit{bag-of-tasks} problems, typically associated with volunteer computing, can also be solved with this approach, by listing each individual task in sequence.

This approach is straight-forward to use with Pando and easiest to combine with other Unix tools. We implemented five applications that show diverse use cases. \textit{Collatz} implements the Collatz Conjecture~\cite{boinc-collatz}, an ongoing BOINC project, to find an integer that results in the largest number of computation steps. Our implementation was compiled from Matlab to JavaScript using the Matjuice compiler~\cite{Foley-Bourgon:2016,matjuice-repository} and then adapted to use a BigNumber library. Other languages with a JavaScript compiler may therefore benefit from Pando without having to implement a distribution strategy. \textit{Raytrace} distributes the rendering of individual frames of a 3D animation and assembles them in an animated gif (Section~\ref{Section:UsageExample}). A similar strategy could be useful to integrate in open source animation tools for artists that do not have access to a rendering farm. \textit{Arxiv} distributes the tagging of interesting papers to a group of collaborators, a form of \textit{crowdprocessing}, by using the browser as a user interface rather than a processing environment. A similar approach could be used to quickly launch an online rescue search using satellite or aerial images in times of disasters. \textit{StreamLender test} performs random executions of StreamLender to find cases where the invariants of the pull-stream protocol are violated. It helped us fix three bugs in corner cases that were not found with manually written tests and then scale up the testing strategy to perform millions of executions quickly without finding errors, increasing confidence that our implementation is correct. \textit{Machine learning agent} searches for the optimal learning rate, an hyperparameter, that helps an autonomous agent in a simulated environment quickly learn sequences of steps that result in rewards. This approach could be beneficial to train deep neural networks in browsers. In this particular example, the training phase is interactive: the user can see the behaviour of the agent as it is learning and early-abort a particular hyper-parameter case if the agent fails to learn, a form a \textit{hybrid human-machine learning collaboration}. \textit{Image processing} blurs the images from the open satellite dataset~\cite{roy2014landsat}. We have implemented multiple versions of this application: this version uses an \textit{http} server to distribute the images and receive the results through http requests. In contrast to the two other versions of Section~\ref{Section:Stubborn}, the data transfer between a Worker and the http server is \textit{synchronous}: a worker processing function will not return a correct result until the output image has been fully transmitted to the server which guarantees that the output image will be received before the output will be produced by Pando.

\subsection{Synchronous Parallel Search}

The structure of \textit{blockchains} in crypto-currencies such as Bitcoin~\cite{nakamoto2008bitcoin} mandates a \textit{synchronous parallel search} organization: all miners compete to find a random value, or \textit{nonce}, such that the hash of the nonce and the block of transactions combined is inferior to a difficulty threshold, itself controlling the probability of finding a nonce. Once a valid nonce has been found, the list of blocks is extended, and all miners start working on the next block.

In the case of Bitcoin, there is no upper bound on the amount of computational power required to mine the next block because the difficulty is automatically adjusted such that the time between each successful block is roughly ten minutes. The increasing difficulty, and therefore computational requirements to mine a new block, makes it increasingly costly for malicious actors to generate a fork of the chain of blocks at arbitrary places, preserving the integrity of the longest chain of blocks. This results in a global consensus on the history of transactions.

\begin{figure}[htbp]
    \begin{center}
    \includegraphics[width=0.23\textwidth]{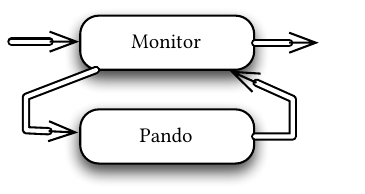}
    \end{center}
\begin{tabular}{llll}
    \textbf{App.}  & \textbf{Inputs}  & \textbf{Monitor}  & \textbf{Pando}   \\
\hline
Crypto-curr.         & Blocks        & Block + Range & Mine nonce            
\end{tabular}
    \caption{\label{Figure:SynchronousParallelSearch} Synchronous parallel search dataflow and example.}
\end{figure}

A synchronous parallel search introduces a \textit{feedback loop} in the flow of data, as illustrated in Figure~\ref{Figure:SynchronousParallelSearch}, because the next input to process is determined by the last valid result obtained. In our implementation, a monitor therefore lazily provides \textit{mining attempts} to Pando, including the current block and a range of integers to test. It generates as many as there are participating workers. Each worker tests all integers in the range and answers either with a valid nonce or a failure and then requests a new mining attempt. The monitor keeps providing new mining attempts until a valid nonce is found and then moves on to the next block. In this example, both the list of blocks and the computational requirements are potentially infinite, making a lazy streaming approach quite natural.

A more efficient implementation would need to relax the ordering constraint to ensure a valid nonce is reported as soon as possible. Otherwise a valid nonce might be held back by other uncompleted work units in front. Adding this support requires only a local change in Pando by adding an option to use a different version of StreamLender that returns \textit{unordered} results.

Moreover, Bitcoin miners nowadays use dedicated hardware that is several orders of magnitude faster than the performance that can be achieved with an equivalent implementation executing in JavaScript. There is therefore limited practicality in mining Bitcoins in browsers, even with the gains obtained by parallelizing the task. Nonetheless, \textit{proof-of-work} algorithms have been designed to work better on regular CPUs~\cite{mukhopadhyay2016brief}. There may therefore be potential applications in mining those emerging crypto-currencies with Pando to support charities and fund open source software.

\subsection{Stubborn Processing with Failure-Prone External Data Distribution}
\label{Section:Stubborn}


 In addition to the http version of Section~\ref{Section:Pipeline}, We implemented two additional versions of distributed blurring of the Landsat-8 open satellite dataset~\cite{roy2014landsat}: one distributing the data with the DAT protocol~\cite{datprotocol}, itself accessible in the Beaker browser~\cite{beaker}, a fork of Chromium~\cite{chromium}, and another that uses  WebTorrent~\cite{webtorrent} running in browsers that support WebRTC.

\begin{figure}[htbp]
    \begin{center}
    \includegraphics[width=0.30\textwidth]{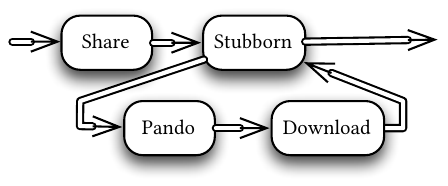}
    \end{center}
    \begin{tabular}{llll}
    \textbf{App.}  & \textbf{Inputs}  & \textbf{Share/Down.}  & \textbf{Pando}   \\
\hline
Img proc.        & Landsat-8 imgs       & DAT protocol & Blur filter     \\   
Img proc.        & Landsat-8 imgs       & WebTorrent protocol & Blur filter     \\       
\end{tabular}
    \caption{\label{Figure:StubbornProcessing} Stubborn processing with external data distribution dataflow and example.}
\end{figure}

In both cases, managing data outside of Pando introduces an additional failure mode due to the \textit{asynchronous} transmission of results: it is possible to receive a successful result but the worker may still crash before the results' data have been fully downloaded. To address the issue, our application outputs a result only after a successful download. Otherwise, the input is resubmitted for computation. The monitoring to implement that feedback loop has been factored into our new \textit{stubborn} pull-stream module~\cite{pull-stubborn} which can be combined with sharing and downloading modules that are specific to a particular protocol, as illustrated in Figure~\ref{Figure:StubbornProcessing}.

This use of Pando could be especially appropriate in cases where there is a growing availability of open datasets combined with limited funding and resources available to process them, as is the case for many citizen initiatives.

\section{Evaluation}
\label{Section:Evaluation}
Our focus in developing Pando has been to easily tap into the computing power of personal devices already owned by the general public. The collective performance of personal devices has previously been shown to be significant both when considering the collection of devices owned by individuals and the aggregate performance of mobile devices of co-workers~\cite{herrera18webassembly,lavoie2019pvc-cf}. The design of Pando has also been shown to scale up to at least a thousand browsers when combined with a fat-tree overlay~\cite{lavoie2019genet} but had not yet been tested on wide-area network deployments. 

In this section, and in complement to the previous results, we compare the performance of Pando on a local area network (LAN) with two additional deployment scenario: a France-wide state-of-the-art computing grid, Grid5000~\cite{grid5000}, connected over a virtual private network (VPN) that is similar to a large organization computing infrastructure, and a wide-area network (WAN) deployment with computing devices distributed throughout Europe on PlanetLab EU~\cite{planet-lab.eu} that is similar to a deployment on the devices of a distributed volunteer community. The throughput results for all three scenario are detailed in Table~\ref{Table:Throughput}: they show that the additional communication latency of the VPN and WAN cases could be hidden by sending multiple inputs at the same time to volunteering devices. Using Pando on compute-bound tasks therefore results in net throughput benefits when using multiple devices in parallel, whether on a LAN, a VPN, or a WAN. In the rest of this section, we detail our experiment settings for all three scenarios, the results obtained, and interesting findings that come from comparing the three scenario together.  To the best of our knowledge, it is the first time an evaluation for a volunteer computing tool has compared those three scales together.

\subsection{Common Settings}

We used all applications of Section~\ref{Section:Applications} except \textit{Arxiv} because the actual "processing" in the Arxiv case is performed by a volunteer rather than the device. All applications are compute-bound, as is typical of volunteer computing. We measured the computation duration and the number of items processed in each Worker over a five minute period, from which we derived the throughput. This diminished the impact of the variability of the computing time between inputs. We also checked that the total of all devices corresponded to the throughput observed at the output of Pando.

The implementation of applications is similar to that used in previous experiments~\cite{lavoie2019pvc-cf}, the only major difference is that the image used for \textit{raytracing} was smaller to avoid a limitation on the size of individual WebRTC messages in the simple-peer~\cite{simple-peer-repo} library we use for managing WebRTC connections. The consequence is that throughput results, in this evaluation, shall be larger for the same devices, running the same browser, on the same network.

Of the three versions of \textit{photo-batch-processing} we implemented, we used the http version, rather than the DAT or the WebTorrent versions. The DAT version can only execute in the Beaker browser~\cite{beaker} because it is the only browser that supports the protocol and its security model requires an explicit confirmation by the user to enable results to be transmitted back, making the test automation cumbersome. The WebTorrent version was not always reliable and sometimes took multiple minutes to establish a connection most probably because the connection of a new node in the underlying WebRTC-based distributed hash table was slow and not always successful. However, choosing the http version meant that the http server that serves files was not accessible from outside a LAN or VPN, we therefore do not provide throughput results on the WAN case. Nonetheless, once peer-to-peer solutions for exchanging files become mature enough, the \textit{image-processing} example shall be easy to adapt to take advantage of their capabilities.

We used Pando version 0.17.14~\cite{pando-repository} with the version of application examples in Pando's handbook~\cite{pando-handbook} at commit \texttt{c5247923}. 

\subsection{LAN: Personal Devices}

We selected a diverse set of devices from our own personal collection, similar to previous experiments on personal devices~\cite{lavoie2019pvc-cf} but omitting the slowest devices and using a more recent version of Pando and applications. We used one iPhone SE (2 cores 1.85 Ghz ARMv8 64-bit), released in 2016, executing iOS 12.1, and Safari. For laptops, we evaluated: (1) a Macbook Air mid-2011 (2 cores i7 1.8 Ghz x86 64-bit) executing MacOS 10.13.6 and Firefox 66.0.5 64-bit; (2) the Novena~\cite{novena}, a linux laptop based on a Freescale iMX6 CPU (4 cores 1.2 Ghz ARMv7 32-bit) produced in a small batch in 2015, executing Debian Linux 8, and Firefox 60.3.0esr 32-bit; (3) an Asus Windows laptop based on a Pentium N3540 (4 cores 2.16 Ghz x86 64-bit) processor executing Windows 10 version 1803 and Firefox 66.0.5 64-bit; and (4) a Macbook Pro 2016 (4 cores i5 2.9 Ghz x86 64-bit) executing MacOS 10.14.1 and Firefox 63.0.1 64-bit. These devices represent a wide variety of CPU and OS choices, as well as a computing performance. We favoured the use of close versions of Firefox on laptops for consistency so the experiments would focus on the variations on CPU speed and because it is generally the fastest on numerical benchmarks~\cite{herrera18webassembly}. We also used the minimum number of cores that provided close to the maximum performance, shown between brackets in Table~\ref{Table:Throughput}; using more cores typically did not significantly increase the total throughput.

The MacBook Air was connected to the other personal devices through a Wifi network. We used a \texttt{batch-size} of 2, effectively enabling one input to be transferred while the other is processed.

\subsection{VPN: Grid5000 Nodes}

We selected one node for each of the 8 participating Grid5000 clusters, themselves distributed between major cities in France along the INRIA network. Each cluster has multiple models, each with a unique name that facilitates selecting a particular model. We list them by model name (ex: \textit{dahu}) followed by the cluster site where they are hosted (ex: \textit{grenoble}), as well as their technical characteristics. They all use different versions of Debian Linux 4.9.x 64-bit and as a browser, Chrome version 73.0.3683.121, through the Electron 5.0.1 environment. 

The nodes were acquired between 2011 and 2018: the oldest is \textit{uvb.sophia} and the most recent is \textit{dahu.grenoble}. Each group of nodes comprises between 15 and 72 nodes. Each node has 2 Intel Xeon CPUs with different model: \textit{uvb.sophia} uses an Intel Xeon X5670 with 6 cores/CPU, while \textit{dahu.grenoble} uses an Intel Xeon Gold 6130 with 16 cores/CPU. The nodes have varying amounts of RAM from 32GB for \textit{petitprince.luxembourg} to 256 GB RAM for \textit{chetemy.lille}. All nodes are connected through 10Gbps ethernet, except for \textit{uvb.sophia} who are connected with 1 Gbps ethernet.

We measured the performance on a single core on a single node per cluster. The results should scale linearly with additional nodes but less than linearly when using more than one core per node, as previous experiments have shown that there is increasing contention for CPU resources when the number of cores used in parallel is increased~\cite{lavoie2019genet}. The Master process of Pando was executing on one core of the MacBook Air 2011, mentioned in the personal devices experiment and the connections between the Master process and the remote devices were made using the WebSocket protocol. The MacBook Air was itself connected to the Internet through the Wifi network of INRIA and to the Grid5000 nodes through a VPN access. We used a \texttt{batch-size} of 2, effectively enabling one input to be transferred while the other is being processed.

\subsection{WAN: PlanetLab EU Nodes}

We selected seven nodes among the PlanetLab EU nodes that are still working and used one core per node. For each node, we used Chrome version 69.0.3497.128 through the Electron 4.1.3 environment. 

Each node has a single Intel CPU, the models comprise a Westmere (\textit{ple42.planet-lab.eu}), a Core 2 Duo (\textit{planet2.elte.hu}), and variations of Xeon (all others). All the nodes have 512MB of RAM, are running Fedora Core Linux version 25 with a 4.8, 4.11, or 4.13 Linux kernel. All nodes are connected through 10 Gpbs ethernet.

We measured the performance on a single core on a single node per cluster. Similar to the VPN experiment, the Master process of Pando was executing on one core of the MacBook Air 2011. However, the connections between the Master process and the remote devices were made using the WebRTC protocol. The MacBook Air was itself connected to the Internet through the Wifi network of INRIA. We used a \texttt{batch-size} of 4, effectively enabling up to three inputs to be transferred while the last is being processed.

\subsection{Analysis}

We highlight here interesting insights from the results of Table~\ref{Table:Throughput}.

\textit{Pando can take advantage of computing devices, whether available on a LAN, a VPN, or a WAN.} We could use the same tool to execute the applications in parallel on personal devices, on a state-of-the-art grid infrastructure, or a distributed set of devices connected to the Internet. In all cases, there was a performance benefit in using all those devices in parallel that improves significantly on the performance that would have been obtained otherwise on a single personal device. To the best of our knowledge, Pando is the first tool for volunteer computing that provides such a level of flexibility. That flexibility, for example, enables leveraging the fastest computing devices available with a minimum of effort: in our experiments, these were the Grid5000 nodes.

\textit{The throughput impact of network latency can be minimized for computation-bound applications, if large enough batches of inputs are used.} For the LAN and VPN experiments, we used input batches of size 2 and for the PlanetLab experiments, we used input batches of 4. These were sufficiently large to compensate for the transmission delay of inputs, even in the case of image-processing where 168kb images were sent for processing through a different channel. Obviously, those results hold only as long the ratio between computation time and data transfer time is sufficiently large. Nonetheless, it shows that for application for which this holds, the option of sending inputs in batches is sufficient to hide the network latency.

\textit{A single core from personal devices of 2016 sometimes provide higher throughput than older servers.} On Collatz, the iPhone SE outperforms the \textit{uvb.sophia} from Grid5000 and almost all PlanetLab server nodes. This is true in more cases when comparing the throughput of a single core on the MBPro 2016 with the performance of a few Grid5000 nodes and many PlanetLab nodes. It therefore means that, sometimes, it may be better to leverage many personal devices than relying on older server nodes.

\textit{The choice of browser sometimes can have dramatic effect on throughput.} The iPhone SE outperforms a single core on the MacBook Pro by 3.3x because Safari performs optimizations that Firefox does not, even if in previous studies Firefox was found to be better in general on numerical computations~\cite{herrera18webassembly}. When using the browser as an execution environment, it is therefore important to try all available browsers to find the best for a specific application.

\textit{2-5 cores on recent personal devices can outperform the fastest server core.} It therefore means that asking 2-5 friends with recent smartphones or laptops, such as the iPhone SE or the Macbook Pro 2016, to participate with Pando can replace renting a high-end server core in remote data centres. While this seems rather impractical if the devices are powered by their battery, the use of portable solar panels can remove the problem during sunny days. 

The previous experiments therefore show that using Pando, a user can leverage spare computing capacity either in local or remote personal devices, that batching inputs is sufficient to hide network latency, and that the computing power available in personal devices is quite significant, even compared to state-of-the-art server infrastructure. 

\section{Related Work}
\label{Section:RelatedWork}

The idea of using idle workstations for distributed computing was first published in 1982~\cite{shoch1982worm} and was then explored in the 90s, 2000s, and 2010s under the umbrella of \textit{desktop grid}~\cite{fedak:hal-00757056, fedak:tel-01158462}. In parallel, \textit{volunteer computing} developed~\cite{sarmenta2001volunteer,anderson2019boinc} to support high-profile research with the personal desktop computers and fast internet connections that were spreading into households.

Individuals nowadays collectively own more computing power, through their personal devices such as desktops, laptops, tablets, phones, etc., than any organization ever did. While there has been work in extending volunteer computing to leverage mobile devices~\cite{tapparello2015volunteer,pramanik2017economical}, the recent \textit{personal} volunteer computing approach~\cite{lavoie2019pvc-cf} is the first to focus on creating \textit{personal tools} for \textit{personal projects} of programmers of the \textit{general public} to seamlessly tap into the computing power of the \textit{personal devices} they, and their \textit{personal social network}, already own.

To the best of our knowledge, Pando is the first tool explicitly designed for the purpose of \textit{personal} volunteer computing.  In this section, we provide more detail on the \textit{declarative concurrency} work it was inspired from and other systems that share similar technology choices. While Pando shares some technology choices with previous platforms, \textit{it combines them for different aims}.

\subsection{Declarative Concurrency}

Declarative concurrency has been studied in the context of dataflow programming, with languages such as Lucid~\cite{lucid} and Oz~\cite{smolka1995oz}. In the Oz language, the declarative programming model can be used directly to implement concurrent modules~\cite[Chapter~4]{van2004concepts}; it is based on using single-assignment variables that enable multiple threads to implicitly synchronize on the availability of data, on top of which higher-level abstractions such as streams can be built. The declarative concurrency paradigm has also been experienced by a large number of programmers and researchers through the popular MapReduce~\cite{dean2008mapreduce} framework and Unix pipeline programming~\cite{Kernighan1983}.  In effect, Pando implements the \textit{map} operation of MapReduce; the other filtering and reduction phases can be performed locally, if necessary, by chaining with other Unix tools, e.g. \texttt{grep} and \texttt{awk}.

JavaScript, as many other mainstream programming languages, has not yet integrated features that make declarative concurrency widely accessible and easy, with good declarative concurrency primitives. We therefore instead based our design and implementation on the pull-stream design pattern (Section~\ref{Section:PullStream}).

As far as we know, we are the first to develop and document systematic abstractions for volunteer computing using the declarative concurrent paradigm.

\subsection{Stream Processing}

\textit{Stream processing} has been widely adopted as a programming model for  \textit{scalable distributed stream processing}~\cite{cherniack2003scalable}, for general purpose programming on CPUs~\cite{gummaraju2005streamCPU}, for distributed GPU programming~\cite{yamagiwa2007design}, and for Web-based peer-to-peer computing based on the WebRTC~\cite{webrtc}, WebSockets~\cite{websocket}, and ZeroMQ~\cite{zeromq} protocols. Those platforms are programmed using \textit{dataflow graphs of computation} that combine multiple operators and complex data flows. They then ensure an efficient and reliable execution on different targeted execution environments. This level of expressivity is not necessary for many personal projects and applications (Section~\ref{Section:Applications}). To support our applications with a lower level of implementation complexity and make our design easier to port to other programming environments, Pando therefore concentrates on distributing the computation that is applied in a single stage of the streaming pipeline with the \textit{map} operation. Everything else is performed locally by leveraging other tools.

\subsection{Browser-Based Volunteer Computing}

Fabisiak et al.~\cite{fabisiak2017browser} have surveyed more than 45 different browser-based volunteer computing systems developed over more than two decades. They grouped the publications in three generations, that followed the evolution of Web technologies: the first generation~\cite{cappello1997javelin,alexandrov1997superweb,baratloo1999charlotte,Sarmenta1998,finkel1999distriblets,nisan1998globally} was based on Java applets; the second generation~\cite{boldrin2007distributed,klein2007unwitting,merelo2008asynchronous,berry2009distributed} used JavaScript instead but was somewhat limited by its performance; and the third generation~\cite{ryza2010mrjs,duda2012distributed,dkebski2013comcutejs,cushing2013distributed,langhans2013crowdsourcing,martinez2015capataz,meeds2015mlitb,macwilliam2013crowdcl} fully emerged once performance issues were solved in multiple ways: JavaScript became competitive with C~\cite{Khan:2014}, WebWorkers~\cite{webworkers}, that did not interrupt the main thread, were introduced, and new technologies, such as WebCL~\cite{webcl}, were proposed to increase the performance beyond what is possible on a single thread of execution on the CPU. 

We further sub-divide Fabisiak and al.'s third generation into an explicit fourth~\cite{kuhara2014peer,leclerc2016space} that incorporates the latest communication technologies, such as WebSocket~\cite{websocket} and WebRTC~\cite{webrtc}, because they make fault-tolerance easier. Pando could be grouped with the fourth generation of systems and, as far as we know, is the first to leverage WebRTC for the explicit goal of volunteer computing. However, the key difference of Pando is in our focus on the \textit{personal} aspects of volunteer computing~\cite{lavoie2019pvc-cf} that led to specific design principles (DPs of Section~\ref{Section:DesignPrinciples}) with the following concrete impacts on its programming model, deployment strategy, and implementation.

Of the systems that have generic \textit{programming models}, many focus on \textit{batch-processing}~\cite{boldrin2007distributed,klein2007unwitting,konishi2007rabc,dkebski2013comcutejs,cushing2013distributed,kuhara2014peer} as typically happens in high-profile long-running applications, sometimes reusing, in the browser, the MapReduce programming model that has been successful in data centers~\cite{berry2009distributed,grigorik2009mapreduce,ryza2010mrjs,langhans2013crowdsourcing,meeds2015mlitb}. In contrast, by using a streaming model, Pando enables different and more personal applications by supporting \textit{infinite streams} and \textit{feedback loops}. This simplifies the combination of Pando with existing Unix tools and other programming environments (DP5).

While some general purpose projects aim to \textit{deploy} new \textit{global platforms}~\cite{alexandrov1997superweb,cappello1997javelin,nisan1998globally,Sarmenta1998,baratloo1999charlotte,konishi2007rabc,dkebski2013comcutejs,langhans2013crowdsourcing,wilkinson2014qmachine,abidi2015towards}, sometimes on \textit{clouds}~\cite{cushing2013distributed,leclerc2016space}, we have chosen to prioritize local deployments for personal uses. Pando also supports cloud platforms, if necessary for connectivity, but our common use cases do not require them. Moreover, by having a deployment that is \textit{specific} to a single user and project (DP1), the implementation is simplified. That removes the need for solutions such as: (1) access restrictions in the form of \textit{random URLs} to segregate the computations of different concurrent users~\cite{wilkinson2014qmachine}, (2) \textit{brokers/dispatchers/bridges} to organize the tasks submitted~\cite{alexandrov1997superweb,cappello1997javelin,konishi2007rabc,dkebski2013comcutejs,langhans2013crowdsourcing,abidi2015towards}, (3) dynamic management of \textit{managers}~\cite{baratloo1999charlotte}, and (4) \textit{advocates}~\cite{Sarmenta1998} to represent clients in the server.

Many \textit{implementations} are organized around a database~\cite{boldrin2007distributed,konishi2007rabc,berry2009distributed,ryza2010mrjs,dkebski2013comcutejs,cushing2013distributed,chorazyk2017volunteer}. Pando's implementation instead encapsulates concurrency aspects in the StreamLender abstraction, removing the need for a database library.  Other implementations are organized around a request-response API based on HTTP~\cite{klein2007unwitting,konishi2007rabc,merelo2008asynchronous,berry2009distributed,ryza2010mrjs,duda2012distributed,dkebski2013comcutejs,cushing2013distributed,langhans2013crowdsourcing,martinez2015capataz,chorazyk2017volunteer}, to distribute inputs and collect results. Instead, and similar to newer projects~\cite{kuhara2014peer,leclerc2016space}, Pando communicates through WebRTC and WebSocket. In our case, the heartbeat mechanism of both protocols enabled our design to encapsulate the fault-tolerance strategy within StreamLender. These simplifications in turn hopefully makes it more likely that other programmers will adapt the design for embedding in other applications or to reimplement as standalone tools for different programming environments.

\subsection{Peer-to-Peer Computing}

\textit{Peer-to-peer computing}, in which participating devices provide resources and help coordinate the services that are used, has a rich literature~\cite{rowstron2001pastry,stoica2001chord,maymounkov2002kademlia,therning2005jalapeno,abdennadher2005towards,nandy2005thesis,harrison2008thesis,kim2009thesis,kim2014scalable,wilson2015architecture,rosen2016thesis}. However,  the server-centric model of Web technologies has historically limited the development of peer-to-peer Web platforms and applications. The recent introduction of WebRTC~\cite{webrtc} removed that limitation which lead to the creation of many new ones~\cite{Johnston:2012,Werner2013,Nurminen2013,Vogt2013,cyclon-webrtc,Hu:2017,dias2018browser}. 

Of all previously mentioned systems, the closest to Pando is \textit{browserCloud.js}~\cite{dias2018browser} in its aim to provide a computation platform powered by the devices of participants. However, Pando's implementation approach is quite different and simpler because a deployment is restricted to a single client, its overlay organization need not make workers communicate with one another, it does not require maintenance when not in use for specific tasks, and removes the need for a discovery algorithm by instead relying on existing social media platforms. In our view, these differences come from a difference in application context. Using BrowserCloud.js's approach, and that of other peer-to-peer systems, is better to create \textit{globally-shared self-sustaining platforms}. Ours is better to quickly obtain a working \textit{personal tool} when a dependency on other tools and platforms is acceptable.

\section{Conclusion}
\label{Section:Conclusion}

In this paper, we presented the design of Pando, a new and first tool for personal volunteer computing that enables a dynamically varying number of failure-prone personal devices contributed by volunteers to parallelize the application of a function on a stream of values using the devices' browsers. In doing so, we have explained how the declarative concurrent model made its programming simple and how the pull-stream design pattern was used to decompose its implementation in reusable modules. We then provided more detail about the properties and implementation of the new StreamLender abstraction that performs the core coordination work within Pando, which, by virtue of being independent of particular communication protocols or input-output libraries, should be easy to reimplement in many other programming environments. We followed with a presentation of a wide variety of novel applications organized along different dataflow patterns that showed Pando was useful on a wide number of existing and emerging use cases. We completed with an evaluation of Pando's benefits in a real-world setting and showed throughput speedups on the previous applications on a local network with personal devices, on a virtual private network spanning France with state-of-the-art server nodes, and a wide area network spanning Europe with older server nodes. The ease and flexibility in deploying Pando shall enable a larger number of programmers to leverage the computing capabilities of personal devices available both locally and remotely. Moreover, our results suggest that the competitive performance of personal devices makes them serious alternative in aggregate for some compute intensive tasks.

%
%

\begin{landscape}
\begin{table}[htbp]
\begin{tabular}{l r r r r r r r r r r r r}
                                                       & \textbf{Collatz}  &            & \textbf{Crypto-}   &           & \textbf{StreamLender-} &      & \textbf{Raytrace} &             &  \textbf{Image-} &           & \textbf{MLAgent-} &  \\
                                                       &                           &            & \textbf{Mining}   &            & \textbf{Testing} &         & \textbf{} &           & \textbf{Process.} &          & \textbf{Training} & \\
                                                        & \textit{Bignum/s} & \textbf{\%}   & \textit{Hashes/s} &  \textbf{\%} & \textit{Tests/s} &\textbf{\%}  & \textit{Frames/s}  & \textbf{\%} & \textit{Images/s} & \textbf{\%} & \textit{Steps/s} & \textbf{\%}  \\
\hline

\hline

\textbf{LAN: Personal Devices \textit{(cores)}}               &  2209.65  &  100.0  &  378,672  &  100.0  &  3603.70   &  100.0  &  18.94  &  100.0    &  0.71  &  100.0  &  484.90  &  100.0  \\
Novena \textit{(2)}                                                          &    121.85  &      5.5  &    16,185  &      4.3   &    142.84  &       4.0  &   0.66   &      3.5   &  0.04  &      5.3   &    51.74  &    10.7 \\
 Asus Laptop  \textit{(3)}                                                &     490.45  &    22.2  &    59,895 &     15.8   &    622.64  &     17.3 &   3.63   &    19.1    &  0.10  &    13.3  &   112.59  &    23.2   \\
\textit{MBAir 2011} \textit{(1)}                                        &     215.58  &      9.8  &    58,693 &     15.5   &    526.82   &     14.6 &   2.94  &     15.5   &  0.06  &      9.0  &     68.81  &    14.2 \\
iPhone SE \textit{(1)}                                                    &      336.18 &     15.2 &     42,720 &     11.3   &    509.64   &     14.1 &   2.90  &     15.3   &  0.33  &    45.9  &     60.24  &    12.4 \\
MBPro 2016 \textit{(2)}                                                 &    1045.58  &    47.3 &   201,178 &     53.1   &  1801.76   &     50.0 &   8.81 &      46.6   &  0.19  &    26.5 &    191.51 &     39.5 \\

\hline
\textbf{VPN: Grid5000 Nodes  \textit{(cores)}}       &   3823.51    &    100.0    & 1,534,102   &     100.0     &  7559.93     &    100.0     &    16.38    &    100.0    &    2.73     &        100.0     &   1323.44   &  100.0  \\
dahu.grenoble \textit{(1)}                            &      642.04   &    16.8      &  230,061     &     15.0        &     1341.77   &      17.7   &       3.12      &   19.0      &     0.44     &        16.1     &  219.18      &  16.6 \\
chetemy.lille                \textit{(1)}                 &      524.71   &    13.7      &  206,195     &     13.5        &      975.58    &      12.9   &       2.04     &    12.5      &    0.37      &       13.6      &  167.03      &   12.6 \\
petitprince.luxembourg   \textit{(1)}             &      261.36   &      6.8       &  136,189    &       8.9        &       631.83   &        8.4   &      1.47      &      9.0       &   0.27      &         9.7      &   124.00     &     9.4 \\
nova.lyon                \textit{(1)}                     &      521.35   &    13.7      &   199,901    &     13.0        &       982.16   &       13.0   &     1.95       &   11.9       &   0.34       &      12.4       &  164.57     &    12.4 \\
grisou.nancy            \textit{(1)}                    &      541.53   &    14.2      &   216,932    &      14.1       &      1026.26  &       13.6   &      2.17      &    13.2      &   0.36      &       13.1       &  176.12     &    13.3 \\
ecotype.nantes            \textit{(1)}                &      479.07   &    12.5       &  187,668    &      12.2       &        939.07  &       12.4   &      1.86      &   11.4       &    0.33      &      12.1       &   162.25     &    12.3 \\
paravance.rennes            \textit{(1)}            &     535.72    &  14.0        &  215,096    &       14.0       &     1021.99   &       13.5  &       2.19      &   13.4      &     0.35      &      12.8       &  176.41     &     13.3 \\
uvb.sophia             \textit{(1)}                      &     317.73    &     8.3        &  142,061    &        9.3        &      641.26    &        8.5   &      1.57      &    9.6       &     0.28      &      10.2       &   133.88     &     10.1 \\

\hline
\textbf{WAN: PlanetLab EU Nodes \textit{(cores)}}  &  1845.52     &   100.0     &  717,485     &   100.0       &  3985.04     &   100.0    &        4.75      &   100.0     &      ---        &       ---          &    714.38    &  100.0 \\
cse-yellow.cse.chalmers.se  \textit{(1)}                      &    470.49     &     25.5     &  162,173     &     22.6       &     996.89    &    25.0     &       0.74       &     15.5     &     ---         &      ---           &    148.85    &   20.8  \\
mars.planetlab.haw-hamburg.de  \textit{(1)}              &    225.38     &     12.2     &    93,189    &       13.0      &     428.30     &   10.7     &       0.64       &      13.6     &     ---        &      ---           &    78.66     &  11.0  \\
ple42.planet-lab.eu  \textit{(1)}                                   &    210.15     &     11.4     &    82,297     &       11.5      &     444.35     &   11.2      &      0.54       &    11.3      &      ---        &     ---            &     81.17    &  11.4 \\
onelab2.pl.sophia.inria.fr  \textit{(1)}                           &    201.43     &     10.9     &    95,609    &        13.3     &     459.66      &  11.5       &     0.68       &    14.3       &   ---          &    ---             &    83.57     & 11.7  \\
planet2.elte.hu \textit{(1)}                                            &   216.42      &     11.7      &  85,927     &       12.0      &     505.04     &   12.7      &      0.73       &    15.4      &    ---          &    ---             &     99.75     & 14.0  \\
planet4.cs.huji.ac.il \textit{(1)}                                     &    298.42      &     16.2     &  112,363   &     15.6        &     651.54     &   16.4      &      0.77       &      16.1     &    ---         &    ---             &    119.62    & 16.7   \\
ple1.cesnet.cz  \textit{(1)}                                           &   223.22      &      12.1    &  85,927      &         12.0    &     499.27     &   12.5      &       0.65      &     13.8      &    ---         &    ---            &     102.76    & 14.4   \\

\end{tabular}
\caption{\label{Table:Throughput} Average throughput for CPU-bound streaming applications on combinations of devices. In all tests, the \textit{MBA 2011} uses one core to execute Pando's Master process, which does not participate in computations, hence only a single core remains for the Personal Devices tests.}
\end{table}
\end{landscape}

\bibliography{ErickLavoie2}

\end{document}